\documentclass[12pt,epsf]{article}
\usepackage{amssymb,amsmath}
\usepackage{graphicx}

\newcommand{\be}{\begin{equation}}
\newcommand{\ee}{\end{equation}}
\newcommand{\bea}{\begin{eqnarray}}
\newcommand{\eea}{\end{eqnarray}}
\newcommand{\beas}{\begin{eqnarray*}}
\newcommand{\eeas}{\end{eqnarray*}}
\newcommand{\ba}{\begin{array}}
\newcommand{\ea}{\end{array}}

\newcommand{\nbox}{{\,\lower0.9pt\vbox{\hrule \hbox{\vrule height 0.2 cm \hskip 0.19 cm \vrule height 0.2 cm}\hrule}\,}}

\def\href#1#2{#2}

\begin{document}
\begin{titlepage}
\hfill
\vbox{
    \halign{#\hfil         \cr
           } 
      }  
\vspace*{20mm}
\begin{center}
{\Large \bf Black Hole Dynamics From Atmospheric Science}

\vspace*{15mm}
\vspace*{1mm}

Mark Van Raamsdonk

\vspace*{1cm}

{Department of Physics and Astronomy,
University of British Columbia\\
6224 Agricultural Road,
Vancouver, B.C., V6T 1W9, Canada}

\vspace*{1cm}
\end{center}

\begin{abstract}

In this note, we derive (to third order in derivatives of the fluid velocity) a 2+1 dimensional theory of fluid dynamics that governs the evolution of generic long-wavelength perturbations of a black brane or large black hole in four-dimensional gravity with negative cosmological constant, applying a systematic procedure developed recently by Bhattacharyya, Hubeny, Minwalla, and Rangamani. In the regime of validity of the fluid-dynamical description, the black-brane evolution will generically correspond to a turbulent flow. Turbulence in 2+1 dimensions has been well studied analytically, numerically, experimentally, and observationally as it provides a first approximation to the large scale dynamics of planetary atmospheres. These studies reveal dramatic differences between fluid flows in 2+1 and 3+1 dimensions, suggesting that the dynamics of perturbed four and five dimensional large AdS black holes may be qualitatively different. However, further investigation is required to understand whether these qualitative differences exist in the regime of fluid dynamics relevant to black hole dynamics.

\end{abstract}

\end{titlepage}

\vskip 1cm

\section{Introduction}

A particularly fascinating aspect of gauge theory / gravity duality is that the high-temperature deconfined phases of gauge theories are mapped via the correspondence to black hole or black brane geometries on the gravity side \cite{witten}. This enables analytic calculations of equilibrium properties of strongly-coupled gauge theory plasmas, such as the deconfinement temperature or the high-temperature equation of state, via a classical analysis of the corresponding geometries.

In the past several years, the connection between gauge theory plasmas and black geometries has also been exploited in the near-equilibrium regime. On the gauge theory side, it is known that long-wavelength fluctuations about the equilibrium state at high temperatures should be described effectively by fluid dynamics equations. Specifically, near-equilibrium configurations may be characterized by a local temperature and a local fluid velocity $u^{\mu}$, in terms of which the stress tensor of the theory can be written as a derivative expansion,
\be
\label{deriv}
T^{\mu \nu} = p(T)\eta^{\mu \nu} + (\varepsilon(T)+p(T))u^\mu u^\nu + {\cal O}(\partial u, \partial T) \; .
\ee
The fluid dynamics equations are then simply the local conservation equations for this stress tensor
\be
\partial_\mu T^{\mu \nu} = 0 \; .
\label{cons}
\ee
The information that distinguishes different field theories in this regime is the set of coefficients appearing in the derivative expansion of the stress tensor, namely the energy density and pressure at zeroth order, the shear viscosity and bulk viscosity at first order, etc... .

While the energy density and pressure are equilibrium quantities, which may be read off from the dual Euclidean black-brane geometry, the viscosities and all higher order coefficients are only relevant for time-dependent processes, and their computation requires a real-time analysis. The first calculation of viscosity appeared in \cite{pss}, where the authors were able to determine the viscosity of maximally supersymmetric gauge theory plasma using a Kubo-type formula, i.e. by studying the low-frequency limit of real-time correlation functions for the stress tensor. Following this, there have been a host of works exploring the fluid dynamics description of high temperature gauge theories with gravity duals (see \cite{ss} for a recent review), motivated in part by the fact that the high-temperature behavior of these theories appears to be qualitatively similar to real QCD. Indeed, this AdS/CFT analysis currently provides the best understanding of certain properties of the quark-gluon plasma produced experimentally in heavy-ion collisions at RHIC.

Recently, Bhattacharyya, Hubeny, Minwalla, and Rangamani \cite{bhmr} have developed a more direct approach for determining the effective fluid dynamical description of a near equilibrium gauge theory plasma with a weakly curved gravity description. This approach provides a systematic way to determine the coefficients in the derivative expansion of the stress-energy tensor to an arbitrary order, and the explicit gravity solution dual to a given solution of the fluid dynamics equations.

The idea of \cite{bhmr} is simply to start with the black brane solution dual to an equilibrium plasma with uniform temperature $T$ and velocity $\vec{\beta}$, allow the temperature and velocity parameters in the metric to be arbitrary slowly-varying functions of the field theory directions, and try to add corrections to this metric order by order in the derivatives of temperature and velocity so that the corrected metric is a valid solution to Einstein's equations. This perturbative procedure works subject to certain constraints on the temperature and velocity fields; these constraints turn out to be exactly the conservation relations (\ref{cons}) for a particular choice of coefficients in the derivative expansion (\ref{deriv}) of the stress-energy tensor. In \cite{bhmr} this procedure was carried out in the case of five-dimensional gravity to determine the complete set of fluid dynamics coefficients up to second order in the derivative expansion. The result for the viscosity matched previous computations, and a subset of the second order coefficients they compute are reproduced using correlation function methods in \cite{brsss}, which appeared simultaneously.

\subsection{Results}

It is worth emphasizing that the calculations in \cite{bhmr} are purely within the context of Einstein gravity with a negative cosmological constant, and do not assume in any way the correctness of gauge theory / gravity duality or any connection to a dual field theory. Thus, the results are also interesting from the perspective of better understanding classical gravitational physics. To reiterate, they say that generic long-wavelength perturbations around a black brane solution in AdS (or, as we will discuss below, a large AdS black hole) are described by fluid dynamics equations. Of course, nonlinear fluid dynamics is extremely difficult to study in general, but decades of analytical, numerical, and experimental research have provided us with a good qualitative understanding and some general quantitative laws for the behavior of fluids in various regimes. We can therefore hope to apply what is known about fluid dynamics to learn new qualitative and quantitative lessons about the {\it generic} behavior of gravity in this nonlinear regime.

To apply these ideas to the most familiar case of 3+1 dimensional gravity, a first step is to generalize the work of \cite{bhmr} (which focused on 3+1 fluid dynamics and therefore 4+1 dimensional gravity) to one lower dimension. This is the main technical goal  of the present paper.

Before giving our results, it will be useful to define exactly what we mean by the temperature field and the velocity field associated with a given field theory (or more generally, with a given stress tensor). For a given point $x$ in the fluid, there will be some velocity $\beta^i(x)$, such that an observer at $x$ moving with this velocity will observe $T^{0i}(x)=0$. We define the proper velocity for this point to be $u^{\mu}(x) = (1, \vec{\beta}(x))/(1-\beta^2(x))^{1/2}$, and we define the temperature for this point to be the temperature observed by the comoving observer (the temperature related by the equilibrium equation of state to the energy density $T^{00}(x)$ as measured in the comoving frame). This definition of velocity may be expressed covariantly as
\[
u_\mu ( T^{\mu \nu} + u^\nu u_\alpha T^{\mu \alpha}) = 0
\]
or equivalently $u_\mu \tilde{T}^{\mu \nu} = 0$ where $\tilde{T}$ represents all terms in $T$ involving derivatives.

With these definitions of velocity and temperature, our result is that the evolution of long-wavelength perturbations to the uniform black-brane solution of 3+1 dimensional gravity with negative cosmological constant is described by the conservation equations (\ref{cons}) where the stress energy tensor to second order in the derivative expansion is
\bea
\label{stress}
T^{\mu \nu} &=& {1 \over 2} \left({4 \pi T \over 3}\right)^3(\eta^{\mu \nu} + 3 u^\mu u^\nu) - \left({4 \pi T \over 3}\right)^2 \sigma^{\mu \nu} \\
 && + {1 \over 18}\left({4 \pi T \over 3}\right)\left[(\sqrt{3} \pi - 9\ln(3)+18) \Sigma_1^{\mu \nu} + 2( \sqrt{3} \pi  - 9 \ln(3)) \Sigma_2^{\mu \nu} \right] \nonumber
\eea
The terms beyond leading order are traceless symmetric tensors which vanish when contracted with $u_\mu$. Defining the operator
\be
\label{proj1}
P^{\mu \nu} = \eta^{\mu \nu} + u^{\mu} u^{\nu}
\ee
which projects vectors in the directions orthogonal to $u$ and
\be
\label{proj2}
\Pi^{\mu \nu}_{\alpha \beta} = {1 \over 2} P^{\mu}_\alpha P^{\nu}_\beta +  {1 \over 2} P^{\mu}_\beta P^{\nu}_\alpha - {1 \over 2} P^{\mu \nu} P_{\alpha \beta}
\ee
which project tensors into traceless symmetric tensors orthogonal to $u^{\mu}$, we have
\bea
\sigma^{\mu \nu} &=& \Pi^{\mu \nu}_{\alpha \beta} \left(  \partial^{\alpha} u^{\beta}  \right) \cr
\omega^{\mu \nu} &=& {1 \over 2} P^\mu_\alpha P^\nu_\beta  \left(  \partial^{\alpha} u^{\beta} -  \partial^{\beta} u^{\alpha} \right)\cr
\Sigma_1^{\mu \nu} &=& \Pi^{\mu \nu}_{\alpha \beta}({\cal D}\sigma^{\alpha \beta}+{1 \over 2} \sigma^{\alpha \beta} \partial_\lambda u^\lambda) \cr
\Sigma_2^{\mu \nu} &=& \Pi^{\mu \nu}_{\alpha \beta}( \sigma_\lambda {}^{\alpha} \omega^{\beta \lambda} )
\eea
where ${\cal D} \equiv u^\alpha \partial_\alpha$. From the expression above, it is straightforward to check that the ratio of viscosity ($-1/2$ times the coefficient of $\sigma$) to entropy density (which works out to $1/T$ times the coefficient of $u^\mu u^\nu$) comes out to the universal \cite{buchel,kss1,kss2} value of $1/(4 \pi)$, first shown in this case by \cite{herzog}.

As reviewed in \cite{brsss, logan}, the AdS isometries (or equivalently, the conformal symmetry of the dual fluid) require that the fluid stress tensor is traceless and covariant under a Weyl transformation of the boundary metric (i.e. the metric of the space on which the dual fluid lives, which we take to be 2+1 dimensional Minkowski space). The two structures $\Sigma_1$ and $\Sigma_2$ turn out to be the only traceless $u$-orthogonal Weyl-covariant structures at two-derivative order relevant for a flat-space theory. In higher dimensions, there are four such structures \cite{brsss,logan}, but two of these (the ones associated with parameters $\lambda_1$ and $\lambda_3$ in the notation of \cite{brsss}) vanish identically for 2+1 dimensional fluid dynamics. In the notation of \cite{brsss}, our results for the second order coefficients are
\beas
\tau_\pi &=& {1 \over 24 \pi T} (\sqrt{3} \pi - 9\ln(3)+18) \cr
\lambda_2 &=& {4 \pi T \over 27}( \sqrt{3} \pi  - 9\ln(3))
\eeas
The result for $\tau_\pi$ agrees exactly with the calculations of \cite{no2,no1}, who computed this using real-time correlation functions, while our result for $\lambda_2$ is new.

In addition to our expression for the stress tensor, which through the conservation equations gives the equations for the effective fluid dynamics to third order in the derivative expansion, the other main technical result is the equation (\ref{final}), which gives the explicit metric dual to an arbitrary solution of the fluid dynamics equations, valid to second order in derivatives.

\subsection{Possible qualitative differences between four and five dimensional black hole evolution.}

The derivative expansion we have presented should be a good description when the length scale $L$ of perturbations is much larger than the inverse temperature
\be
\label{condition}
LT \gg 1 \; .
\ee
In this limit, the most important terms are the leading terms in the derivative expansion, and these are the terms conventionally included in studies of fluid dynamics. While our calculations explicitly refer to perturbations around a uniform black-brane solution, dual to an unbounded fluid, the same fluid dynamics on a sphere of radius $R \gg {1 \over T}$ should describe the evolution of perturbations around a large AdS-Schwarzschild black hole (i.e. one whose radius is large compared with the length scale of AdS curvature). \footnote{This follows from a simple AdS/CFT argument, since a thermal CFT on Minkowski space dual to the black-brane arises from the same CFT at finite temperature on a sphere in the limit of large sphere volume and fixed temperature. The CFT on sphere with $R \gg 1/T$ is dual to a large AdS black hole, and this CFT should have locally the same long-wavelength physics as the flat-space one, since correlation lengths in this limit will be much smaller than the sphere size.}

At the level of the equations of motion, our results look very similar to the ones obtained in \cite{bhmr} for 4+1 dimensional gravity. Despite the similarity of the equations, the relation to fluid dynamics suggests that there could be dramatic differences between the physics of large AdS black holes in four versus five dimensions. To understand these, we first note that the regime $LT \gg 1$ where our derivative expansion is valid is also the regime of large Reynolds number (where the viscous terms are small relative to the leading order terms), so we expect that generic evolution of the long-wavelength perturbations about the black brane will correspond to fluid flows in the turbulent regime. In 3+1 dimensions, turbulent flows are characterized by an ``energy cascade'' in which large scale eddies give rise to smaller scale eddies, tending towards configurations of increasing disorder and transferring energy down to scales where viscosity becomes important and energy is dissipated (converted to heat). Thus, 3+1 dimensional fluids are rather efficient at dissipating initial fluctuations, regardless of how small the viscosity is. For a basic review of fluid dynamics and turbulence, see \cite{landau, frisch}.

In contrast, turbulent flows in 2+1 dimensional fluids are characterized by an ``inverse cascade,'' in which smaller scale eddies merge into large scale eddies, eventually creating large persistent vortical structures from which the energy is dissipated only very slowly, since the dissipative terms are relatively unimportant at large scales (see section 9.7 of \cite{frisch}, and also \cite{vH}). In unbounded systems the late-time behavior for generic initial conditions is believed to be a ``dilute gas of vortices'' while in bounded systems (such as fluid flow on a sphere), the flow evolves to a single vortex or a small number of vortices with size comparable to that of the system. These two-dimensional phenomena are relevant in nature, since two-dimensional turbulence provides a first approximation to large scale motions in planetary atmospheres, and also to the evolution of large-scale oceanic currents (examples of the persistent vortical structures include intense tropical storms, the polar vortices, and the Great Red Spot of Jupiter).

These qualitative differences between two and three dimensional fluid dynamics suggest that the evolution of generically perturbed AdS black holes may exhibit significant qualitative differences between two and three dimensions. Taken at face value, the results of the previous paragraphs would suggest that black brane and large black-hole configurations in five-dimensional gravity with negative cosmological constant equilibrate much faster than those in four dimensions, and furthermore, that the late-time behavior for generic perturbations around a four-dimensional black brane or large $AdS_4$ black hole will be characterized by the formation of long-lived large-scale vortical structures, related to the more familiar ones observed in planetary atmospheres .

However, there is an important caveat.\footnote{I would like to thank Veronika Hubeny and Mukund Rangamani for motivating me to understand this better.} The fluid dynamical results we refer to are valid in the context of Navier-Stokes equations for non-relativistic fluids which have $p \ll \epsilon$ (in analytical and numerical studies, the fluids are usually also taken to be incompressible). On the other hand, the fluids we discuss have pressure and energy density of the same order of magnitude; in other words, they are microscopically relativistic. For such fluids, the equations of motion differ from Navier-Stokes equations even in the limit where all the macroscopic velocities are small. It is plausible that the qualitative differences between two and three dimensional fluid dynamics are not just restricted to the Navier-Stokes regime. However, before making any specific conclusions about about differences between four and five dimensional black hole evolution, it will be important to understand whether the qualitative differences we have noted (or other differences) exist in the regime where the fluids are microscopically relativistic. A recent discussion highlighting some features of two-dimensional turbulence in relativistic fluids has appeared in \cite{romatschke}.

It is interesting to ask whether any of this discussion is applicable to astrophysical black holes (which should behave like small black holes in AdS). Unfortunately, the present analysis doesn't shed any light on this question, since for ordinary Schwarzschild black holes (or small AdS black holes), the length scale associated with temperature is just the size of the black hole. Thus, the condition (\ref{condition}) for the validity of the hydrodynamic approximation cannot be satisfied; equivalently, all higher order terms in the expansion of the stress-energy tensor are equally relevant.

\subsection{String theory applications}

While we have emphasized that our results are apply completely within the context of four-dimensional gravity, there are also important applications within the context of string theory using the AdS /CFT correspondence. According to the correspondence, the fluid dynamical theory we have worked out provides the effective description of long-wavelength perturbations around the thermal equilibrium state for any 2+1 dimensional conformal field theory with a weakly curved gravity dual, provided that Einstein gravity with negative cosmological constant is a consistent truncation of the corresponding gravity theory. The simplest example is the $SO(8)$ conformal field theory that provides the worldvolume theory of a large number of M2-branes in M-theory. For this theory (and other theories) there are locally conserved charges in addition to the energy and momentum. The fluid description we have found applies to neutral fluctuations about the uncharged equilibrium state, but it should be straightforward to determine a more complete effective theory which allows for nonzero charge densities by applying a similar analysis starting with a charged black-brane solution.

\subsection{Outline}

The remainder of this paper is organized as follows. In section 2, we present the perturbative calculation of the metric for a generic non-singular perturbation of the 3+1 dimensional black-brane solution to second order in the derivative expansion, and derive the associated constraints on velocity and temperature fields. In section 3, we use our expression for the metric to calculate the boundary stress tensor, using the prescription of \cite{bk}. This stress tensor is interpreted as the stress tensor of the dual fluid, and we find that the constraint equations found in section 2 coincide with the conservation equations for this stress tensor, expanded to second order in derivatives.

\section{Construction of the perturbed solution}

In this section, we apply the methods of \cite{bhmr} to perturbatively construct a solution of Einstein gravity with negative cosmological constant dual to an arbitrary solution of particular 2+1 dimensional fluid dynamics equations that we will also construct in the process.

\subsection{Unperturbed metric}

We begin with Einstein's equations in 3+1 dimensions with negative cosmological constant
\be
\label{einstein}
R_{IJ} - {1 \over 2} g_{IJ} R - \lambda g_{IJ} = 0 \; ,
\ee
where for convenience, we choose units such that $\lambda = 3$. The uniform black brane solution corresponding to a temperature $T$ is\footnote{These coordinates, introduced in \cite{bhmr} are related to more typical coordinates where the metric is diagonal by a coordinate transformation $v = t + h(r)$, where $h'(r) = 1/(r^2 f(r))$}
\[
ds^2 = 2 dv dr - r^2 f(b r) dv^2 + r^2 d\vec{x}^2
\]
where\footnote{As usual, the relationship between temperature and the parameter b may be determined by demanding that there is no conical singularity in the Euclidean continuation when the Euclidean time direction is chosen to have periodicity $1/T$.}
\[
f(r) = 1 - {1 \over r^3} \qquad \qquad b = {3 \over 4 \pi T} \; .
\]

Boosting this solution to a proper velocity $u^\mu$ along the $\vec{x}$ directions, we obtain the three-parameter family of solutions
\be
\label{zeroth}
ds^2 = -2 u_\mu dx^\mu dr - r^2 f(b r) u_\mu u_\nu dx^\mu dx^\nu + r^2 P_{\mu \nu} dx^\mu dx^\nu
\ee
where $u^\mu$ is the proper velocity
\[
u^0 = {1 \over \sqrt{1 - \vec{\beta}^2}} \qquad \qquad u^i = {\beta^i \over \sqrt{1 - \vec{\beta}^2}}
\]
and $P_{\mu \nu} = \eta_{\mu \nu} + u_\mu u_\nu$ is the projection into the directions orthogonal to $u$, defined so that $u^\mu P_{\mu \nu} = 0 $.

This solution depends on the temperature $T$ and velocity $\vec{\beta}$, so we have a map between equilibrium configurations of a fluid moving at constant velocity and uniform black-brane solutions. We would now like to extend this correspondence to near-equilibrium configurations in which the temperature and velocity of the fluid are allowed to vary in space, evolving in time according to fluid dynamics equations. Thus, we now promote $b$ and $\beta$ to general functions of space and time (assumed to vary slowly on the scale $1/T$) in the metric (\ref{zeroth}), and try to add corrections to this, order by order in derivatives of $b$ and $\beta$, such that the corrected metric is a solution to Einstein's equations (\ref{einstein}). As in \cite{bhmr}, we will find that this is possible as long as $b(x^\mu)$ and $\beta(x^\mu)$ satisfy fluid dynamics equations with particular choices for the pressure, energy density, viscosity, and higher order coefficients.

\subsection{Overview of perturbation theory}

We now review the perturbative procedure developed in \cite{bhmr}. At each order, we begin with a metric $g^{(n-1)}$ depending on $b(x^\mu)$ and $\beta(x^\mu)$ which solves Einstein's equations up to terms involving $n-1$st derivatives of $b$ and $\beta$, but for which terms in Einstein's equations involving $n$th and higher derivatives do not cancel. For $n=1$, the metric is just the one in (\ref{zeroth}) with $b$ and $\beta$ taken to depend on space and time. We then try to add an $n$th order metric $g^{(n)}$ such that the complete metric solves Einstein's equations up to $n$th order in derivatives.

Throughout the perturbative procedure, we work with the gauge choice \cite{bhmr}
\be
\label{gauge}
g_{rr} = 0 \qquad \qquad g_{r \mu } \propto u_\mu \qquad \qquad (g^{(0)})^{\mu \nu} g^{(n>0)}_{\mu \nu} = 2 g^{(n)}_{rv} + {1 \over r^2} g^{(n)}_{ii} = 0
\ee
The metric is further constrained by the requirement that its components
$g_{\mu r}$ and $g_{\mu \nu}$ must transform as a vector field and a tensor field under 2+1 dimensional Lorentz transformations where $u_\mu$ and $b$ are taken to transform as vector and scalar fields. This follows from the covariance of the leading order metric and the covariance of Einstein's equations. The most general covariant metric at $n$th order in derivatives satisfying our gauge condition is
\bea
\label{general}
(ds^2)^{(n)} &=& {k_n \over r^2} u_\mu u_\nu dx^\mu dx^\nu - 2 h_n u_\mu dx^\mu dr -  r^2 h_n P_{\mu \nu} d x^\mu dx^\nu \cr
&&\qquad \qquad - {2 \over r} (j_n)_\nu u_\mu dx^\mu dx^\nu +  r^2 (\alpha_n)_{\mu \nu} dx^\mu dx^\nu
\eea
where we may choose
\[
u_\mu j^\mu =0  \qquad \qquad u_\mu \alpha^{\mu \nu} = 0 \qquad \qquad \eta_{\mu \nu} \alpha^{\mu \nu} = 0 \; .
\]
Here $k_n(r,u_\mu(x),b(x))$, $h_n(r,u_\mu(x),b(x))$, $j_n^\mu(r,u_\mu(x),b(x))$ and $\alpha_n^{\mu \nu}(r,u_\mu(x),b(x))$ are expressions involving a total of $n$ derivatives of $u_\mu(x)$ and $b(x)$, that transform respectively as two scalar fields, a vector field, and a symmetric traceless tensor field when $u_\mu$ and $b$ are taken to transform as vector and scalar fields under 2+1 Lorentz transformations. Note that the only dependence of these functions on the coordinates $x^\mu$ is through the functions $b(x^\mu)$ and $u(x^\mu)$.

In order to determine the functional dependence of $k$, $h$, $j$, and $\alpha$ on $r$, $b$, and $u$, it is enough to solve Einstein's equations at any particular point, so we choose to work at $x^\mu = 0$ and further, take coordinates where $b(0)=1$ and $\beta_i(0)=0$.

In this case, the extra terms (\ref{general}) that we add at $n$th order in perturbation theory reduce to
\[
(ds^2)^{(n)} = {k_n(r) \over r^2} dv^2 + 2 h_n(r) dv dr - r^2 h_n(r) d x^i dx^i + {2 \over r} j_n^i(r) dv dx^i + r^2 \alpha^{ij}_n(r) dx^i dx^j
\]
Using this undetermined expression and the results from lower order, we now take
\[
g = g^{(0)} + \dots g^{(n)},
\]
Taylor expand about $x=0$ to $n$th order in derivatives of $b$ and $\beta$, plug into Einstein's equations, which may be rewritten as
\be
\label{wij}
W_{IJ} \equiv R_{IJ} + 3 g_{IJ} = 0 \; ,
\ee
and try to choose $h_n$, $k_n$, $j^i_n$, and $\alpha^{ij}_n$ so that all terms with $n$ derivatives cancel (terms with less than $n$ derivatives will cancel assuming that we have correctly carried out the perturbative procedure at lower orders). We will find that this is possible so long as $b(x)$ and $\beta(x)$ obey certain constraint equations, which amount to the equations of fluid dynamics with specific coefficients for pressure, energy density, viscosity, etc...

\subsection{General structure of perturbation theory}

Carrying out the procedure above, we find that at $n$th order in perturbation theory, the functions $h_n$, $k_n$, $j^i_n$ and $\alpha^{ij}_n$ may be determined by the four equations
\bea
\ba{rclcrcl} W^{(n)}_{rr} & = & 0 & \qquad \implies \qquad & {1 \over r^4} {d \over dr} (r^4 h_n'(r)) &=& S^{(n)}_h(r) \cr
r^4f(r) W^{(n)}_{rr} - W^{(n)}_{ii} & = & 0 & \qquad \implies \qquad & {d \over dr} (-{2 \over r} k_n(r)+(1-4r^3 ) h_n(r))) &=&  S^{(n)}_k(r)\cr
W^{(n)}_{ri} &=& 0 & \qquad \implies \qquad & {r \over 2} {d \over dr}({1 \over r^2} {d \over dr} \vec{j}_n(r)) &=& \vec{S}^{(n)}_j(r)\cr
W^{(n)}_{ij}-{1 \over 2} \delta_{ij} W^{(n)}_{ii} &=& 0 & \qquad \implies \qquad & {d \over dr}(-{1 \over 2} r^4 f(r) {d \over dr} \alpha_n^{ij}(r)) &=& {\bf S}^{(n)}_{\alpha}(r) \ea
\label{master}
\eea
where $W^{(n)}_{IJ}$ denotes all $n$ derivative terms in $W_{IJ}$ (defined in (\ref{wij})) and the equations on the right come from dividing these terms into those that arise from $g^{(n)}$ and those that arise from lower order terms in the metric.

The general solution to these equations is given by the specific solution
\bea
h_n(r) &=& \int_r^\infty  {dx \over x^4} \int_x^\infty dy y^4 S^{(n)}_h(y) \cr
k_n(r) &=& -{r \over 2} \int_r^\infty dx S^{(n)}_k(x) + {r \over 2} (1 - 4 r^3) \int_r^\infty  {dx \over x^4} \int_x^\infty dy y^4 S^{(n)}_h(y) \cr
\vec{j}_n(r) &=& 2 \int_r^\infty dx x^2 \int_x^\infty  {dy \over y} S^{(n)}_h(y) \cr
{\bf \alpha_n} &=& \int_r^\infty  {2 dx \over x^4 f(x)} \int_1^\infty dy {\bf S}^{(n)}_\alpha(y)
\label{solutions}
\eea
plus the general solution to the homogeneous (source-free) equations,
\bea
h_n(r) &=& s_n + {t_n \over r^3}  \cr
k_n(r) &=& u_n r + {t_n \over 2 r^2} - 2 r^4 s_n \cr
\vec{j}_n(r) &=& \vec{a}_n + {\vec{b}_n \over 3} r^3  \cr
{\bf \alpha_n} &=& {\bf c}_n + { 1 \over 3} {\bf d}_n \ln(r^2-{1 \over r}) \; .
\label{homo}
\eea
With our choice for the the specific solution, it turns out that we can/must set all coefficients in the homogeneous solution to zero. First, we must set $s_n$, $\vec{b}_n$, ${\bf c}_n$ and ${\bf d}_n$ to zero in order to get a nonsingular solution that preserves the original asymptotics. The constant $t_n$ may be set to zero by a coordinate transformation $r \to r + t_n/(2r^2)$ that preserves our gauge choice. The coefficients $u_n$ and $\vec{a}_n$ may be adjusted freely by redefinitions of $b$ and $\vec{\beta}$. But we chose to fix the definitions of $b$ and $\beta$ by demanding that $u_\mu T^{\mu \nu}=0$. For any non-zero $u_n$ or $\vec{a}_n$, this condition is violated, so we must demand that these constants vanish. Then the $b$ and $\beta$ that appear in the metric are directly related to the inverse temperature and velocity of the fluid.

Once we have solved for $h$, $k$, $j$, and $\alpha$ at a given order, the remaining four equations $W_{rv} = W_{vv} = W_{vi} = 0$ give rise to constraints on the set of $n$-derivative expressions built from temperature and velocity. These constraints turn out to be precisely the fluid dynamics equations
\[
\partial_\mu T^{\mu \nu}
\]
Taylor expanded to $n$th order in derivatives about the point $x=0$. The constraints at order $n$ involve the stress tensor at order $(n-1)$, which can be computed as the boundary stress tensor of the metric obtained at the previous order in perturbation theory.

Given the solutions for $h$, $k$, $j$, and $\alpha$ we can write the metric at order $n$ in the derivative expansion. From this, we can calculate the boundary stress-energy tensor (which we interpret as the stress-energy tensor of the dual fluid) at order $n$, using the prescription of \cite{bk} reviewed in section 3.

In the next two subsections, we provide explicit details of the perturbative calculation at first order and second order in derivatives.

\subsection{First order}

At first order, the source terms in equation (\ref{master}) are
\beas
S^{(1)}_h &=& 0 \cr
S^{(1)}_k &=& -4 r \partial_i \beta_i \cr
\vec{S}^{(1)}_j &=& -{1 \over r} \partial_v \vec{\beta} \cr
{\bf S}^{(1)}_\alpha &=& 2r {\bf \sigma}
\eeas
where
\[
\sigma_{ij} = {1 \over 2}(\partial_i \beta_j + \partial_j \beta_i - {1 \over 2} \delta_{ij} \partial_k \beta_k) \; .
\]
With these sources, the metric components at first order may be calculated using (\ref{solutions}) and give
\beas
h_1(r) & =& 0 \cr
k_1(r) &=& r^3 \partial_i \beta_i \cr
\vec{j}_1(r) &=& r^2 \partial_v \vec{\beta} \cr
{\bf \alpha_1} &=& 2 {\bf \sigma} F(r) \sim {\bf \sigma} ({2 \over r}- {2 \over 3 r^3} + {\cal O}(r^{-4}))
\eeas
where
\be
\label{deff}
F(r) = -{\sqrt{3} \over 3} {\rm Tan}^{-1}\left( {\sqrt{3} \over 3}(2r+1) \right) + {1 \over 2} \ln(1 + {1 \over r} + {1 \over r^2}) + {\sqrt{3} \pi \over 6} \; .
\ee
With these assignments, we find that the remaining Einstein equations are satisfied if and only if
\bea
\partial_v b &=& {1 \over 2} \partial_i \beta_i \cr
\partial_i b &=& \partial_v \beta
\label{const1}
\eea
These are the terms obtained at first order in derivatives in the conservation equation $\partial_\mu T^{\mu \nu} = 0$, where we need only keep the non-derivative terms in the expression (\ref{stress}) for $T$.

\subsection{Second order}

To discuss the results at second order, it is useful as in \cite{bhmr} to catalogue all the scalar, vector, and traceless symmetric tensor terms built from $b$ and $\beta$ at two-derivative order, since these are the expressions that will appear in the second order metric. We find:
\beas
\ba{lll} {\rm Scalar} & {\rm Vector} & {\rm Tensor} \cr
{\cal S}_1 = \partial_v^2 b & {\cal V}_{1i} = \partial_v \partial_i b &  {\cal T}_{1ij} =  \partial_i \partial_j b - {1 \over 2} \delta_{ij} \partial^2 b \cr
{\cal S}_2 = \partial_i^2 b & {\cal V}_{2i} = \partial_v^2 \beta_i &  {\cal T}_{2ij} =  \partial_v \sigma_{ij} \cr
{\cal S}_3 =\partial_v \partial_i \beta_i &  {\cal V}_{3i} = \partial_i \partial_j \beta_j &  {\cal T}_{3ij} =  \partial_v \beta_i \partial_v \beta_j -{1 \over 2} \delta_{ij} (\partial_v \beta_k)^2  \cr
{\cal S}_4 = \partial_v \beta_i \partial_v \beta_i & {\cal V}_{4i} = \partial^2 \beta_i &  {\cal T}_{4ij} =  \partial_k \beta_i \partial_k \beta_j -{1 \over 2} \delta_{ij} (\partial_k \beta_l)^2  \cr
{\cal S}_5 = (\partial_i \beta_i)^2 & {\cal V}_{5i} = \partial_v \beta_i \partial_j \beta_j &  {\cal T}_{5ij} =  \partial_i \beta_k \partial_j \beta_k -{1 \over 2} \delta_{ij} (\partial_k \beta_l)^2\cr
{\cal S}_6 = (\epsilon_{ij} \partial_i \beta_j)^2 & {\cal V}_{6i} = \partial_v \beta_j \partial_i \beta_j &\cr
{\cal S}_7 = \sigma_{ij} \sigma_{ij} & {\cal V}_{7i} = \partial_v \beta_j \partial_j \beta_i &\cr
\ea
\eeas
Here, we have excluded pseudoscalar, pseudovector and pseudotensor terms, since these will not appear in the metric. Also, note that in 2+1 dimensions, we have
\beas
\sigma_{ij} \partial_k \beta_k &=& {1 \over 2}({\cal T}_{4ij}+ {\cal T}_{5ij}) \cr
\partial_k \beta_{(i} \partial_{j)} \beta_k - {1 \over 2} \delta_{ij} \partial_k \beta_l \partial_l \beta_k &=& {1 \over 2}({\cal T}_{4ij}+ {\cal T}_{5ij})
\eeas
so we do not need to include the structures on the left, which are independent in higher dimensions.

The Einstein equations at second order in derivatives have a solution provided that the following constraints are satisfied
\bea
{\cal S}_1 &=& {1 \over 2} {\cal S}_3 - {1 \over 2} {\cal S}_4 + {1 \over 4} {\cal S}_5 \cr
{\cal S}_2 &=& {\cal S}_3 + {\cal S}_4 - {1 \over 2} {\cal S}_6 + {\cal S}_7\cr
{\cal V}_1 &=& {1 \over 2} {\cal V}_3 + {1 \over 2} {\cal V}_5 -{1 \over 2} {\cal V}_6 \cr
{\cal V}_2 &=& {1 \over 2} {\cal V}_3 + {1 \over 2} {\cal V}_5 -{1 \over 2} {\cal V}_6 - {\cal V}_7 \cr
{\cal T}_1 &=& {\cal T}_2 + {\cal T}_3 + {1 \over 4} {\cal T}_4 + {1 \over 4} {\cal T}_5 \cr
\partial_v \partial_{[i}\beta_{j]} &=& {1 \over 2} \partial_k \beta_k \partial_{[i}\beta_{j]} \; .
\label{const2}
\eea
In addition, the first order constraint (\ref{const1}) must be corrected to include higher order terms
\bea
\partial_v b &=& {1 \over 2} \partial_i \beta_i - {1 \over 3} {\cal S}_7 \cr
\partial_i b &=& \partial_v \beta - {1 \over 3} {\cal V}_4 -{2 \over 3} {\cal V}_5 + {2 \over 3} {\cal V}_6 \; .
\label{const3}
\eea
It may be checked that the four constraints (\ref{const3}) and the nine constraints (\ref{const2}) are exactly the 4+9 equations arising from the equations $\partial_\mu T^{\mu \nu} |_{x=0} = 0$ and $\partial_\alpha \partial_\mu T^{\mu \nu}|_{x=0} =0$ expanded to second order in derivatives.\footnote{In \cite{bhmr}, the temperature field $b$ was decomposed into terms $b^{(n)}$ such that the constraints set term $b^{(n)}$ equal to a sum of $n$ derivative terms involving $\beta$ without any further corrections. Our $b$ corresponds to their $\sum_n b^{(n)}$.}

Taking into account the constraints, we can now write the second order source terms in equation (\ref{master}) in terms of the independent two-derivative expressions and solve for the metric at second order. We have
\bea
S^{(2)}_h &=& -{1 \over 2 r^4} {\cal S}_6 + F_1(r) {\cal S}_7 \cr
S^{(2)}_k &=& 2 {\cal S}_3 +{1 \over 2} {\cal S}_5 -2(1+{1 \over r^3}) {\cal S}_6 + F_2(r) {\cal S}_7\cr
\vec{S}^{(2)}_j &=& {1 \over 2 r^2} {\cal V}_3 - {1 \over 2r^2(1+r+r^2)} {\cal V}_4 - {r^2+r-1 \over 2 r^2 (r^2+r+1)}({\cal V}_6 - {\cal V}_5) \cr
{\bf S}^{(2)}_\alpha &=& F_3(r) ( {\cal T}_2 + {\cal T}_3) + F_4(r) {\cal T}_4 + F_5(r) {\cal T}_5
\label{sources2}
\eea
where
\beas
F_1(r) &=& {2(1+2r) \over  r^2 (r^2 + r + 1)^2}F(r) -{ (r+1)^2 \over r^2 (r^2 + r + 1)^2} \cr
F_2(r) &=& 2F(r){1+3r+4r^2-4r^3-6r^4-4r^5 \over r^2(1+r+r^2)}  + {4r^3+4r^2+2r-1 \over r(1+r+r^2)} \cr
F_3(r) &=& 2r F(r) -{2r(r+1) \over r^2 + r +1 } \cr
F_4(r) &=& -{r \over 2} F(r) - {1 \over 2(r^2+r+1)} \cr
F_5(r) &=& {3 \over 2} r F(r) - {2r^2 + 2r-1 \over 2(r^2+r+1)}
\eeas
and $F(r)$ was defined in (\ref{deff}).

With these sources, the metric components at second order are given by the expressions in (\ref{solutions}). For our purposes of calculating the second order stress tensor, we only need to know the asymptotic behavior of the various functions for large $r$. We find:
\beas
h_2(r) &=& {1 \over r^2} ({1 \over 2} {\cal S}_7 + {1 \over 4} {\cal S}_6) + {\cal O}(r^{-4}) \cr
k_1(r) &=& r^2({\cal S}_7 + {1 \over 2} {\cal S}_6 -{1 \over 4} {\cal S}_5 -{\cal S}_3)+ {\cal O}(r^0)  \cr
\vec{j}_1(r) &=& r(-{1 \over 2}{\cal V}_3 -{1 \over 2} {\cal V}_5 + {1 \over 2} {\cal V}_6) + {\cal O}(r^{-1}) \cr
{\bf \alpha_1} &=& {1 \over 2 r^2}({\cal T}_5 - {\cal T}_4) + {1 \over r^3}\left({1 \over 12}({\sqrt{3} \pi \over 9} - \ln(3)+2)(4{\cal T}_2 + 4{\cal T}_3 - {\cal T}_4 + 3{\cal T}_5) + {1 \over 3}({\cal T}_4 - {\cal T}_5)\right) \cr
&& \qquad + {\cal O}(r^{-4})
\eeas
In all these expressions, the higher order terms do not give any finite contributions to the boundary stress tensor.

\subsection{Covariant form of the metric}

We have now derived complete expressions for the functions $h$, $k$, $j$, and $\alpha$ appearing in the expression (\ref{general}) for the metric corrections at first and second order in derivatives. However, to derive these expressions, we have been working at a particular point $x=0$ with a choice of coordinates where $b(0)=1$ and $\beta_i(0) =0$. To recover the general expression for the metric without these assumptions, we only need to rescale coordinates $r \to br$, $x^{\mu} \to x^\mu/b$ and find Lorentz covariant expressions $k$, $h$, $j_\mu$, and $\alpha_{\mu \nu}$ that reduce to our expressions above in the frame where $\beta_i(0) =0$.

The final result for the metric is
\bea
ds^2 &=& -2 u_\mu dx^\mu dr - r^2 f(b r) u_\mu u_\nu dx^\mu dx^\nu + r^2 P_{\mu \nu} dx^\mu dx^\nu \cr
 && +r \partial_\lambda u^\lambda u_\mu u_\nu dx^\mu dx^\nu - r u^\lambda \partial_\lambda( u_\mu u_\nu) dx^\mu dx^\nu +2 r^2 b F(br) \sigma_{\mu \nu} dx^\mu dx^\nu \cr
 && + {k_2( b r) \over b^2 r^2} u_\mu u_\nu dx^\mu dx^\nu - 2 b^2 h_2(br) u_\mu dx^\mu dr -  r^2 b^2 h_2(br) P_{\mu \nu} d x^\mu dx^\nu \cr
&&\qquad  \qquad \qquad \qquad - {2 \over b r} j_{2\nu}(b r) u_\mu dx^\mu dx^\nu +  b^2 r^2 \alpha^{\mu \nu}_2(br) dx_\mu dx_\nu
\label{final}
\eea
where $k_2$, $h_2$, $j_2$ and $\alpha_2$ are defined in terms of the sources (\ref{sources2}) by (\ref{solutions}), and we replace the expressions ${\cal S}$, ${\cal V}$, and ${\cal T}$ built from derivatives of $\beta$ with the covariant ones given by
\beas
{\cal S}_3 &=& u^{\mu} \partial_\mu \partial_\nu u^{\nu} - {\cal D} u^\mu {\cal D} u^{\mu}  \cr
{\cal S}_4 &=& {\cal D} u_\mu {\cal D} u^\mu \cr
{\cal S}_5 &=& (\partial_\mu u^\mu)^2 \cr
{\cal S}_6 &=& -(\epsilon^{\mu \nu \lambda} u_\mu \partial_\nu u_\lambda)^2 \cr
{\cal S}_7 &=& \sigma^{\mu \nu} \sigma_{\mu \nu} \cr
{\cal V}_{3\mu} &=& P_\mu^\sigma (\partial_\sigma \partial_\nu u^\nu - {\cal D} u_\nu \partial_\sigma u^{\nu})  \cr
{\cal V}_{4\mu} &=& P_\mu^\sigma P^{\rho \nu} \partial_\rho \partial_\nu u_\sigma \cr
{\cal V}_{5\mu} &=& P_\mu^\sigma \partial_\nu u^\nu {\cal D} u_\sigma \cr
{\cal V}_{6\mu} &=& P_\mu^\sigma {\cal D} u_\nu \partial_\sigma u^{\nu} \cr
{\cal T}^{\mu \nu}_2 &=& \Pi^{\mu \nu}_{\alpha \beta} \left( {\cal D} \partial^\alpha u^\beta \right)\cr
{\cal T}^{\mu \nu}_3 &=& \Pi^{\mu \nu}_{\alpha \beta} \left( {\cal D} u^\alpha {\cal D} u^\beta  \right) \cr
{\cal T}^{\mu \nu}_4 &=& \Pi^{\mu \nu}_{\alpha \beta} \left( \partial_\rho u^\alpha  \partial^\rho u^\beta + {\cal D} u^\alpha {\cal D} u^\beta\right) \cr
{\cal T}^{\mu \nu}_5 &=& \Pi^{\mu \nu}_{\alpha \beta} \left(\partial^\alpha u^\rho \partial^\beta u_\rho \right)\cr \; .
\label{covT}
\eeas
Here, $P^{\mu \nu}$  and $\Pi^{\mu \nu}_{\alpha \beta}$ were defined in (\ref{proj1}) and (\ref{proj2})  and ${\cal D} \equiv u^\alpha \partial_\alpha$.

\section{Stress tensor}

The boundary stress-energy tensor (which we associate with the stress-energy tensor of the dual fluid) may be computed using the prescription of \cite{bk}, as
\[
T^{\mu \nu} =  \lim_{r \to \infty} (r^5 (\Theta^{\mu \nu} - \gamma^{\mu \nu} \Theta-2\gamma^{\mu \nu} - G^{\mu \nu}))
\]
Here, $\Theta^{\mu \nu}$ is the extrinsic curvature for the surface of constant $r$, which may be calculated as
\[
\Theta^{I J} = \nabla^{I} v^{J} - v^{I} v_{K} \nabla^K v^{K}
\]
where $v^{K}$ is the vector field of unit vectors normal to the surface of constant $r$, determined by
\[
g_{I \mu} v^I = 0 \qquad \qquad g_{IJ} v^I v^J = 1 \; .
\]
All other tensors are constructed using the boundary metric $\gamma_{\mu \nu}$ induced on the surface at fixed $r$, and $G^{\mu \nu}$ is the Einstein tensor calculated from this metric (with indices raised by $\gamma$).

Using the metric we have derived, it is straightforward to calculate the stress tensor to second order in derivatives. Note that one derivative and two derivative terms in the metric $g_{\mu \nu}$ of order $r$ and $1$ respectively give rise to potential divergences in the stress-energy tensor (scaling as $r^2$ and $r$ respectively), but these all cancel. The only terms contributing to the finite part of the stress tensor are the $1/r$ terms in $g_{\mu \nu}$, which include the zero-derivative terms proportional to $u_\mu u_\nu$ and the one and two-derivative terms in the symmetric traceless spatial terms in the metric (i.e. the $\alpha_{ij}$ terms). The final result for the stress tensor to two-derivative order is given as equation (\ref{stress}). It may be checked that the structures appearing at second order may be written in terms of the covariant two-derivative tensors defined in the previous section as
\beas
\Sigma_1 &=&  {\cal T}_2 + {\cal T}_3 + {1 \over 4}{\cal T}_5 + {1 \over 4}{\cal T}_4\cr
\Sigma_2 &=& {1 \over 4}{\cal T}_5 - {1 \over 4}{\cal T}_4; .
\eeas

\section*{Acknowledgements}

I am grateful to Veronika Hubeny, Mukund Rangamani and especially Shiraz Minwalla for helpful comments. I would also like to thank R. Loganayagam for pointing out typos in the original version of the draft. This work has been supported in part by the Natural Sciences and Engineering Research Council of Canada, the Alfred P. Sloan Foundation, and the Canada Research Chairs programme.


\begin{thebibliography}{999}

\bibitem{witten}
  E.~Witten,
  ``Anti-de Sitter space, thermal phase transition, and confinement in  gauge
  theories,''
  Adv.\ Theor.\ Math.\ Phys.\  {\bf 2}, 505 (1998)
  [arXiv:hep-th/9803131].

\bibitem{bhmr}
  S.~Bhattacharyya, V.~E.~Hubeny, S.~Minwalla and M.~Rangamani,
  ``Nonlinear Fluid Dynamics from Gravity,''
  arXiv:0712.2456 [hep-th].

\bibitem{pss}
  G.~Policastro, D.~T.~Son and A.~O.~Starinets,
   ``The shear viscosity of strongly coupled N = 4 supersymmetric Yang-Mills plasma,''
  Phys.\ Rev.\ Lett.\  {\bf 87}, 081601 (2001)
  [arXiv:hep-th/0104066].

\bibitem{brsss}
  R.~Baier, P.~Romatschke, D.~T.~Son, A.~O.~Starinets and M.~A.~Stephanov,
  ``Relativistic viscous hydrodynamics, conformal invariance, and holography,''
  arXiv:0712.2451 [hep-th].

\bibitem{herzog}
  C.~P.~Herzog,
  JHEP {\bf 0212}, 026 (2002)
  [arXiv:hep-th/0210126].

\bibitem{no2}
  M.~Natsuume and T.~Okamura,
  ``A note on causal hydrodynamics for M-theory branes,''
  arXiv:0801.1797 [hep-th].

\bibitem{no1}
  M.~Natsuume and T.~Okamura,
  ``Causal hydrodynamics of gauge theory plasmas from AdS/CFT duality,''
  arXiv:0712.2916 [hep-th].


\bibitem{logan}
  R.~Loganayagam,
  ``Entropy Current in Conformal Hydrodynamics,''
  arXiv:0801.3701 [hep-th].

\bibitem{landau}
L. D. Landau and E. M. Lifshitz, ``Fluid Mechanics,'' 2nd edition. Pergamon Press, Oxford, 1987

\bibitem{frisch}
U. Frisch, ``Turbulence,'' Cambridge University Press, 1995

\bibitem{vH}
G.J.F. van Heijst,``Self-Organization of Two-Dimensional Flows,''
Nederlands Tijdschrift voor Natuurkunde 59, 321-325 (1993),
http://www.fluid.tue.nl/WDY/2Dturb/ntvn/selforg.html


\bibitem{ss}
  D.~T.~Son and A.~O.~Starinets,
  ``Viscosity, Black Holes, and Quantum Field Theory,''
  Ann.\ Rev.\ Nucl.\ Part.\ Sci.\  {\bf 57}, 95 (2007)
  [arXiv:0704.0240 [hep-th]].

\bibitem{bk}
  V.~Balasubramanian and P.~Kraus,
  ``A stress tensor for anti-de Sitter gravity,''
  Commun.\ Math.\ Phys.\  {\bf 208}, 413 (1999)
  [arXiv:hep-th/9902121].

\bibitem{buchel}
  A.~Buchel and J.~T.~Liu,
  ``Universality of the shear viscosity in supergravity,''
  Phys.\ Rev.\ Lett.\  {\bf 93}, 090602 (2004)
  [arXiv:hep-th/0311175].

\bibitem{kss1}
  P.~Kovtun, D.~T.~Son and A.~O.~Starinets,
  ``Holography and hydrodynamics: Diffusion on stretched horizons,''
  JHEP {\bf 0310}, 064 (2003)
  [arXiv:hep-th/0309213].

\bibitem{kss2}
  P.~Kovtun, D.~T.~Son and A.~O.~Starinets,
  ``Viscosity in strongly interacting quantum field theories from black hole
physics,''
  Phys.\ Rev.\ Lett.\  {\bf 94}, 111601 (2005)
  [arXiv:hep-th/0405231].

\bibitem{romatschke}
  P.~Romatschke,
  ``Fluid turbulence and eddy viscosity in relativistic heavy-ion collisions,''
  arXiv:0710.0016 [nucl-th].




\end{thebibliography}
\end{document}